\documentclass[twocolumn]{aastex701}
\usepackage{amsmath}
\usepackage{xcolor}
\usepackage[utf8x]{inputenc}
\usepackage[T1]{fontenc}
\newcommand{\ang}{\r{A}}
\hypersetup{linkcolor=blue, urlcolor=blue}
\begin{document}
\title{Investigating the Multiwavelength Emission of Binary SMBH Candidate SDSS J095036.75+512838.1}
\author[orcid=0009-0001-3446-5776]{Cassandra Daniele}
\affiliation{Department of Physics, University of Maryland, Baltimore County, 
1000 Hilltop Cir, Baltimore, MD 21250, USA}
\email{cdaniel3@umbc.edu}

\author[0000-0002-1616-1701]{Adi Foord}
\affiliation{Department of Physics, University of Maryland, Baltimore County, 
1000 Hilltop Cir, Baltimore, MD 21250, USA}
\email{foord@umbc.edu}

\author[0000-0001-8557-2822]{Jessie Runnoe}
\affiliation{Department of Physics \& Astronomy, Vanderbilt University, 
1400 18th Ave. South, Nashville, TN 32712, USA}
\affiliation{Department of Life and Physical Sciences, Fisk University, 
1000 17th Avenue N, Nashville, TN 37208, USA}
\email{jessie.c.runnoe@vanderbilt.edu}

\author[0000-0002-3719-940X]{Michael Eracleous}
\affiliation{Department of Astronomy and Astrophysics and Institute for Gravitation and the Cosmos, Penn State University, 
525 Davey Lab, 251 Pollock Road, University Park, PA 16802, USA}
\email{mxe17@psu.edu}

\author[0000-0001-9806-4034]{Niana N. Mohammed}
\affiliation{Department of Astronomy and Astrophysics and Institute for Gravitation and the Cosmos, Penn State University, 
525 Davey Lab, 251 Pollock Road, University Park, PA 16802, USA}
\email{nnm5189@psu.edu}

\begin{abstract}
We present results from a multiwavelength study of the binary AGN candidate SDSS J095036.75+512838.1, which was identified as a candidate binary SMBH system due to its significant and variable broad H$\beta$ line shifts. We use available infrared-to-X-ray data to investigate the system's accretion properties. Analyzing recent Chandra observations, we characterize the X-ray emission and test the binary SMBH hypothesis. We find that the X-ray spectrum is well-modeled as a single AGN, with a 2--10 keV luminosity of $1.3 \times10^{43}$ erg s$^{-1}$. By assembling a comprehensive multiwavelength spectral energy distribution, we further examine the possibility of circumbinary accretion. The infrared—optical emission matches that expected from a single AGN, though we observe an excess in emission between $\sim 9500-19000$ \ang\ and a deficit near $\sim 2600$ \ang. Archival optical spectra also suggest a deficit near $\sim 8500$ \ang. We investigate whether these SED anomalies can be explained by long-term variability or interstellar reddening. We find no evidence of dimming, while reddening can account for the deficit in UV emission. However, the origin of the near-infrared excess remains unclear. Ultimately, the SED of SDSS J095036 is well described by emission from a single, reddened AGN. Future high-resolution infrared and X-ray spectral observations will be crucial for disentangling the origin of the infrared excess and testing the binary SMBH scenario in SDSS J095036.
\end{abstract}
\section{Introduction}
Galaxy mergers are thought to play an important role in galaxy evolution and trigger supermassive black hole (SMBH) growth (\citealt{kauffmann_unified_2000}, \citealt{hopkins_unified_2006}, \citealt{goulding_galaxy_2018}, \citealt{comerford_excess_2024}). Because most massive galaxies host a central SMBH (\citealt{kormendy_coevolution_2013}), their merging inherently results in the pairing of SMBHs within the combined gravitational potential of the new, merged system (\citealt{begelman_massive_1980}). A merging system becomes a binary once the two SMBHs become gravitationally bound, which typically occurs at separations of $\lesssim 10$ parsec (\citealt{dotti_supermassive_2007}, \citealt{bogdanovic_electromagnetic_2022}). As these SMBHs spiral in towards each other, they emit low-frequency gravitational-waves, the predicted dominant component of the stochastic background detected by pulsar timing arrays (PTAs; \citealt{agazie_nanograv_2023}, \citealt{burke-spolaor_astrophysics_2019}, \citealt{burke-spolaor_era_2025}). Binary SMBHs represent the last stage of the inspiral toward coalescence, and studying them electromagnetically provides an observable connection to gravitational-wave events detectable by future space-based laser interferometers (\citealt{colpi_lisa_2024}). 
 
In these later stages of the evolution, stellar hardening is the main driver of energy loss in the system. At this stage, three-body interactions where individual stars scatter off the binary system can strip away orbital energy and angular momentum. Stellar hardening is possible when there are stars within the system's loss cone -- once emptied the binary theoretically can stall (i.e. the ``final parsec problem'', \citealt{begelman_massive_1980},\citealt{milosavljevic_long-term_2003}). Yet many simulations and models now suggest other ways to continue orbital decay (e.g., \citealt{merritt_chaotic_2004}). As the system approaches milli-parsec scale separations, gravitational waves become the dominant mechanism for angular momentum loss. Studying binary candidates can help determine how frequently these systems reach sub-pc scales and merge, and allow us to better understand the physical processes governing their in-spiral.

In order to detect binary systems electromagnetically, at least one SMBH must be actively accreting matter as an active galactic nucleus (AGN). The most conclusive method to confirm a binary is by spatially resolving two individual AGN. Currently, spatially resolving binary SMBHs is severely limited by their close separations and available telescope point spread functions, requiring the resolving power of radio interferometry, such as in the discovery of 0402+379 (\citealt{rodriguez_compact_2006}, \citealt{bansal_constraining_2017}). However, this technique is limited to low redshifts and requires both AGN to be radio-loud. This results in very few detections, as the resolving power can only probe sub-pc separations up to $z=0.1$ (at z=0.1, 1~pc corresponds to an angular separation of $\sim0.5$mas), and radio-loud AGN only accounts for $\sim$15$\%$ of all AGN (e.g, \citealt{burke-spolaor_radio_2011}, \citealt{kellermann_radio-loud_2016}). 

Thus, clear observational evidence for binaries remains limited to a small number of systems (e.g, \citealt{rodriguez_compact_2006}, \citealt{valtonen_massive_2008}). Due to these challenges in directly detecting binary SMBHs, researchers have developed indirect techniques for identification. One of the most common approaches is through photometric variability characterized by periodicity in quasar light curves. This periodicity can be caused by the secondary periodically intercepting the primary SMBH's accretion disk (such as the case of OJ287, see \citealt{valtonen_massive_2008}), accretion via a circumbinary disk  (\citealt{macfadyen_eccentric_2008}, \citealt{roedig_evolution_2012}, \citealt{farris_binary_2014}, \citealt{roedig_observational_2014}, \citealt{lai_circumbinary_2023}), or jet procession caused by the presence of a companion (\citealt{deane_close-pair_2014}, \citealt{tiede_disk-induced_2024}). However, AGN light curves exhibit stochastic variability, or red-noise variability,  which can be mistaken for a periodic signature in the case of a limited baseline (\citealt{vaughan_false_2016}). Systematic searches for periodic variability in quasar light curves—including more recent studies using X-ray variability (\citealt{liu_bat_2020}, \citealt{tubin-arenas_searching_2025})—have identified numerous candidate binary SMBH systems (e.g., \citealt{valtonen_massive_2008}, \citealt{graham_systematic_2015}, \citealt{liu_periodically_2015}, \citealt{charisi_population_2016}, \citealt{guo_spectral_2020}, \citealt{liao_discovery_2021}, \citealt{luo_systematic_2024}). However, the results of these searches have been mixed, with some large-scale surveys recovering few or no compelling candidates, and no sub-milliparsec binary SMBHs through periodicity alone.

A complementary approach is to search for velocity shifts in the broad emission lines of AGN spectra, which can arise from the orbital motion of one SMBH relative to the observer. Systematic spectroscopic searches have identified promising binary SMBH candidates through significant and/or variable broad-line velocity offsets (e.g., \citealt{tsalmantza_systematic_2011}, \citealt{eracleous_large_2012}, \citealt{ju_search_2013}, \citealt{shen_constraining_2013}, \citealt{liu_constraining_2014}, \citealt{graham_systematic_2015}, \citealt{charisi_population_2016}), although these signatures can also be produced by single SMBH phenomena such as disk emitters, outflows, or transient accretion events.
 
One alternative approach uses multi-wavelength observations to search for spectral signatures of accretion from a circumbinary disk, as different emission mechanisms dominate at different energies (\citealt{foord_multi-wavelength_2017}, \citealt{foord_investigating_2022}). In the canonical binary SMBH picture, each black hole possesses its own mini-disk, while the system as a whole is surrounded by a circumbinary disk (e.g. \citealt{roedig_evolution_2012}, \citealt{gultekin_observable_2012}, \citealt{tanaka_electromagnetic_2012}, \citealt{tanaka_electromagnetic_2013}, \citealt{roedig_observational_2014}). In
this scenario there is a gap between these disks, at a radius which would otherwise emit in the optical to UV wavelengths, resulting in decreased emission in these bands. 
 
Whether this picture applies to all binary SMBHs remains uncertain. Early theoretical work suggested that, for very close binaries or when the specific angular momentum of infalling gas is low, mini-disks may fail to form altogether (\citealt{gultekin_observable_2012}; \citealt{tanaka_electromagnetic_2013}; \citealt{gold_accretion_2014}; \citealt{roedig_observational_2014}). More recent simulations, however, indicate that mini-disks can survive to much smaller separations than previously thought, although they become increasingly truncated (\citealt{bowen_relativistic_2017}; \citealt{dascoli_electromagnetic_2018}; \citealt{wang_final_2023}).

When mini-disks are present, binary SMBHs are predicted to exhibit a characteristic optical--UV ``notch'' in their SEDs, the location and depth of which depend on the total mass, mass ratio, and semi-major axis of the system (e.g., \citealt{roedig_evolution_2012}; \citealt{roedig_observational_2014}; \citealt{farris_binary_2014}). At the same time, supersonic gas streams flowing from the circumbinary disk onto the mini-disks are expected to produce shocks that generate a Wien-like excess in the hard X-ray (\citealt{roedig_observational_2014}; \citealt{farris_characteristic_2015}; \citealt{farris_binary_2015}). The additional high-energy emission from these shocks can partially fill in the optical--UV notch, complicating the predicted SED. As a result, the detailed spectral signatures depend sensitively on the binary configuration and accretion flow. Observational evidence for these predictions remains limited: searches for the expected ``notch'' or hard X-ray excess have found no compelling evidence in binary candidates ( \citealt{foord_multi-wavelength_2017}; \citealt{foord_investigating_2022}; \citealt{saade_nustar_2024}), although statistical studies have identified systematic differences between the broadband SEDs of binary SMBH candidates and those of single AGN (\citealt{lusso_nature_2014}).
\subsection{Binary SMBH candidate SDSS J095036.75+512838.1}
With this framework in mind, we present a multi-wavelength analysis of binary SMBH candidate SDSS J095036.75+512838.1 (hereafter J095036) at z=0.2142. J095036 is among 88 $z \lesssim 0.7$ SDSS quasars identified by \cite{eracleous_large_2012} that exhibit broad $H\beta$ line offsets exceeding 1000 km s$^{-1}$.  J095036 was identified as an especially strong candidate due to its systematic and monotonic velocity shifts of several hundred km s$^{-1}$, which are difficult to reconcile with a recoiling black hole scenario, where the broad-line velocity offset would be expected to remain approximately constant over observational timescales comparable to a human lifetime \citep{runnoe_large_2017}. Assuming the binary SMBH hypothesis, \cite{nguyen_emission_2020} compared the observed H$\beta$ emission-line profiles of the \cite{eracleous_large_2012} sample to a large suite of semi-analytic circumbinary disk models using a principal component analysis framework, deriving a mass ratio of $q=0.43 \pm 0.17$. More recently, \cite{mohammed_tantalizing_2026} carried out an in-depth analysis of the broad $H\beta$ lines of J095036 using optical spectra from SDSS, the Palomar Hale 5m Telescope (Pal), the Keck 10m telescope, and the Hobby-Eberly 11m Telescope (HET) spanning 22 years. Their findings further support the binary SMBH scenario, revealing both a significant systematic shift in the $H\beta$ line and only approximately $20\%$ stochastic variability in the integrated flux, ruling out various other plausible theories such as a dust-obscuration scenario. Under the binary hypothesis, they determine a best-fit period of $T=33^{+7}_{-3}$ years, an eccentricity of $e=0.65^{+0.13}_{-0.13}$, and lower limits on the semi-major axis and black hole mass of $a=10^{-2}$ pc (at z=0.2142, $\theta \sim 3\times10^{-3}$ mas) and $M_{BH} \ge 10^7 M_{\odot}$, respectively. \cite{breiding_chandra_2026} analyze Chandra X-ray observations of velocity-offset quasars from the \cite{eracleous_large_2012} sample, including J095036. They compute various mass estimates based on different AGN and host galaxy properties searching for systematic differences that may indicate a binary system. They find no such evidence for the sample as a whole or individual systems. For J095036, they report an optical to X-ray luminosity ratio of $\alpha_{ox}=1.45\pm0.11$, an associated Eddington ratio of $\log \lambda=-0.9\pm0.5$, and mass estimates in the range $M_{BH}=10^{7.3-8.8} M_{\odot}$.

To further evaluate the nature of J095036 and search for evidence supporting or contradicting a binary SMBH scenario, we analyze available X-ray observations from Chandra, as well as archival mid-infrared (Wide-field Infrared Survey Explorer; WISE, \citealt{https://doi.org/10.26131/irsa1}, \citealt{cutri_explanatory_2013}), near-infrared (Two Micron All-Sky Survey; 2MASS, \citealt{skrutskie_two_2006}, \citealt{https://doi.org/10.26131/irsa2}), optical (SDSS, \citealt{york_sloan_2000}), and ultraviolet (Galaxy Evolution Explorer; GALEX, \citealt{martin_galaxy_2005}, available at MAST: \dataset[10.17909/T9H59D]{http://dx.doi.org/10.17909/T9H59D}) data. This broadband dataset allows us to characterize the accretion properties of J095036 and test whether its observed emission is consistent with expectations for a binary SMBH system.

Our paper is organized in the following manner. In Section \ref{sec:Xray} we analyze the X-ray observation of J095036 and evaluate the 0.5-8 keV spectrum for evidence of a binary SMBH system; in Section \ref{sec:SED} we present the multi-wavelength SED and compare the emission to single AGN spectral models; in Section \ref{sec:results} we discuss our results and test for effects of dimming and reddening; and in Section \ref{sec:conclusion} we review our conclusions. We assume a standard $\Lambda$CDM cosmology of  $\Omega_{\Lambda} = 0.7$, $\Omega_M = 0.3$, and $H_0= 70$ km s$^{-1}$ Mpc$^{-1}$.

\section{X-ray Observations} \label{sec:Xray}
\begin{figure*}[t]
    \centering
    \includegraphics[width=0.7\linewidth]{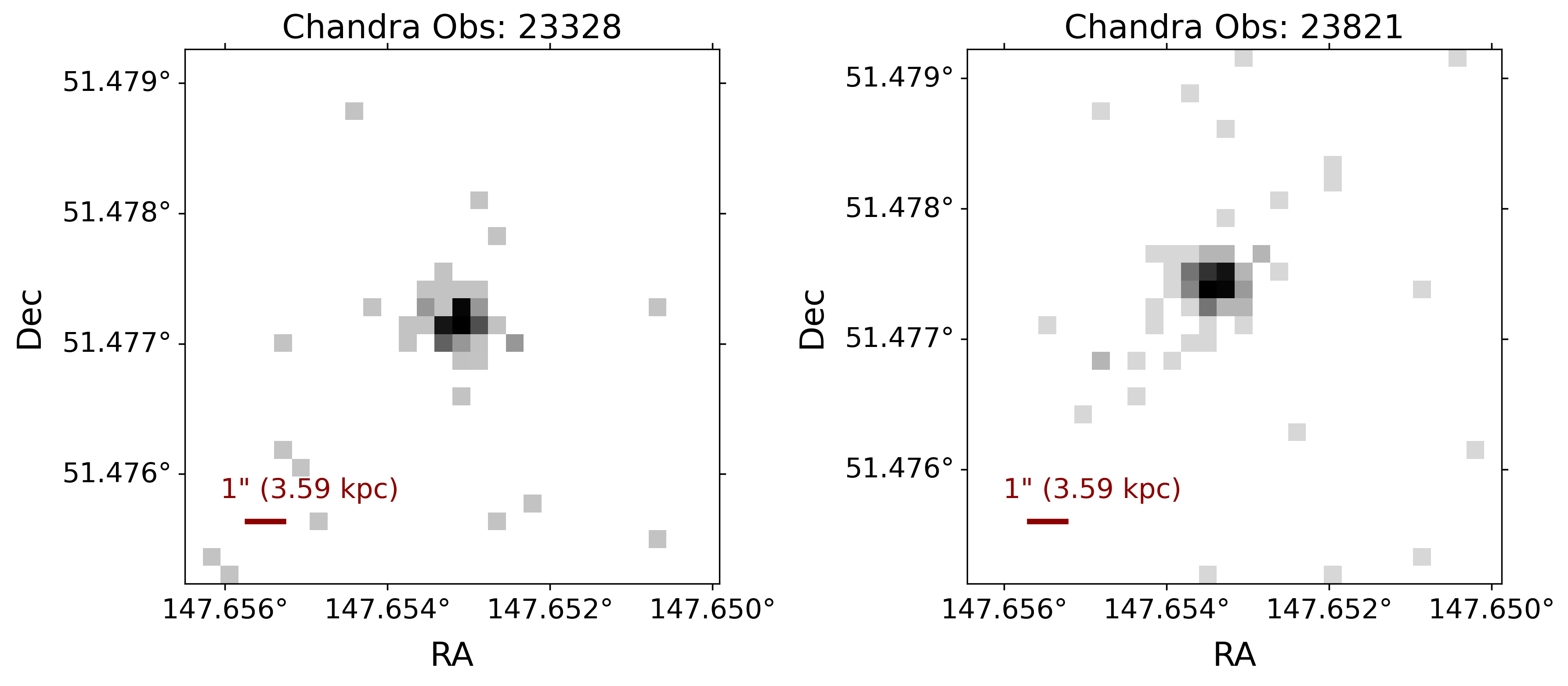}
    \caption{The 0.5 to 10 keV Chandra images of J095036. Observation 23328 (left) and 23821 (right) have 4.99 ks and 12.88 ks of exposure time after running \texttt{deflare}. The total 0.5-10 keV counts within the source region of radius $2 \arcsec$ from the Chandra detection coordinates are 73 and 120 respectively. A $1 \arcsec$ line (3.59 kpc at z=0.2142) is shown for scale. }
    \label{fig:chandra}
\end{figure*}

J095036 was observed twice during Cycle 22 with the Chandra X-ray Observatory (ObsIDs 23328 and 23821; PI: Nyland), using the ACIS-S instrument (S3 aimpoint). The observations were carried out on 2020 November 05 and 2021 January 15 UT for 4.99 ks and 12.89 ks, respectively. J095036 was targeted for X-ray follow-up based on Very Large Array Sky Survey results identifying it as a changing-state quasar, having transitioned from radio-quiet to radio-loud over the past 1–2 decades \citep{nyland_quasars_2020}. We reduce the observations using the Chandra Interactive Analysis of Observations (CIAO) software package (\citealt{fruscione_ciao_2006}) v4.17. Following the standard reprocessing steps, we first run the CIAO command \texttt{chandra\_repro} on the secondary data products.We then correct for background flares by removing time intervals where the count rate deviated by more than $3\sigma$ from the mean level using \texttt{deflare}. We additionally remove low-count-rate time bins at the beginning and end of each observation, applying cuts of $<0.01$ ks. We assess the astrometry by cross-matching Chandra-detected sources with the SDSS DR12 catalog using \texttt{wcs$\_$match}. We require matches within 2\arcsec and a minimum of three counterparts. Neither observation satisfies these criteria and no astrometric correction was applied. 
For both observations, we detect an X-ray source coincident with the SDSS position of J095036, with the X-ray centroid located within $<1\arcsec$ of the optical coordinates. We extract the unbinned source spectra using a circular region with a radius of $2\arcsec$ centered on the X-ray source, and extract background spectra from a source-free annulus with inner and outer radii of $10\arcsec$ and $30\arcsec$, respectively. All spectra are extracted using \texttt{specextract}.
The resulting 0.5–10 keV source counts are 73 and 120 for ObsIDs 23328 and 23821, respectively. 
The Chandra images of J095036 are shown in Figure \ref{fig:chandra}.

\subsection{X-ray Spectral Analysis} \label{sec:xray_spec}
We perform all spectral fitting using XSPEC (\citealt{arnaud_xspec_1996}) version 12.14.1. All errors reported in the following sections are evaluated at the 95\% confidence level and calculated using the XSPEC Monte Carlo Markov Chain tool \texttt{chain}. We first fit each observation individually to assess the accretion state in each epoch, followed by a joint analysis. We fit the 0.5 to 8 keV emission with 3 different models: Model 1, an absorbed redshifted power law (\texttt{phabs$\times$zphabs$\times$zpow}), Model 2, a partially absorbed redshifted power law (\texttt{phabs$\times$(zphabs$\times$zpow$+$zpow})) with power law indices tied, and Model 3, an absorbed redshifted broken power law (\texttt{phabs$\times$zphabs$\times$bknpower}) to test for the presence of two accretion disks by allowing for two different accretion rates. 

For all models, we fix the Galactic hydrogen column density ($N_H$ associated with \texttt{phabs}) to a value of $1.03 \times 10^{20}$ cm$^{-2}$ (calculated using \citealt{bekhti_hi4pi_2016}). We set the extragalactic absorption component ($N_{\rm H,int}$ associated with \texttt{zphabs}) to a value of $1.0 \times 10^{22}$ cm$^{-2}$ and allow it to vary within bounds spanning the Galactic column density as a lower limit and $5.0 \times 10^{24}$ cm$^{-2}$ as an upper limit. For all photon indices ($\Gamma$) associated with \texttt{zpow}, we  restrict the range to 1–3 (see \citealt{ishibashi_x-ray_2010}), with an initial value of 1.8. For Model 3, the second photon index associated with \texttt{bknpower} is initialized at $\Gamma_2 = 2.2$ to provide a realistic but different value from that of $\Gamma_1$ (set to 1.8). Additionally, the break energy of the broken power law is allowed to vary between $0.5$ and $5$ keV, with an initial value of $2.7$ keV chosen near the midpoint of this range.

Given the low number of counts associated with each observation (72 and 119 counts for 23328 and 23821, respectively), we implement the Cash statistic (\citealt{cash_parameter_1979}) to compare models. We quantify whether one model is preferred over another by evaluating if there is a statistically significant improvement, \textit{i.e.} $\Delta C_{stat} > 2.71 \times \Delta N_{fp}$ (where $\Delta N_{fp}$ is the difference in the number of free parameters between the models; \citealt{tozzi_x-ray_2006}, \citealt{brightman_evolution_2012}), corresponding to a 90\% improvement of the fit (\citealt{brightman_compton_2014}) so long as the best-fit model remains physically reasonable (i.e. no values are pegged to the boundaries of the allowed range). 

\begin{deluxetable*}{ccccccccc}
    \tablecaption{Chandra Spectral Fits}
    \tablewidth{0pt}
    \tablehead{
    \colhead{Observation} & \colhead{Model} & \colhead{$N_H$} & \colhead{$\Gamma_1$} &\colhead{ $\Gamma_2$} & \colhead{$E_{break}$} & \colhead{$F_{2 -10 keV}$} & \colhead{$L_{2-10 keV}$} & \colhead{$C_{stat}/dof/N_{fp}$} \\
    (1) & (2) & (3) & (4) & (5) & (6) & (7) & (8) & (9)  }
    \startdata
    23328 & 1 & $\leq 1.0 \times10^{-2} $ & $2.0_{-0.5}^{+0.6}$ & ... & ... & $1.5_{-0.5}^{+0.6}$ & $1.9_{-0.6}^{+0.8}$ & 253.34/509/3 \\
      & 2 & $50_{-30}^{+50}$ & $2.7_{-0.6}^{+0.3}$ & ... & ... & $1.5_{-0.2}^{+1.5}$ & $1.9_{-0.4}^{+1.0}$ & 253.32/508/4\\
      & 3 & $ \leq 1.0 \times10^{-2}$ & $2.6_{-0.9}^{+0.4}{^\dagger}$ & $1.4_{-0.4}^{+0.8}{^\dagger}$ &$2.4_{-1.2}^{+2.1}$ &  $2.0_{-0.9}^{+0.7}$ & $2.3_{-0.8}^{+0.5}$ & 245.54/507/5 \\
      \hline
    23821 & 1 & $ \leq 1.0 \times10^{-2}$ & $2.4_{-0.6}^{+0.5}$ & ... &... & $0.7_{-0.3}^{+0.3}$ & $1.0_{-0.3}^{+0.3}$ &278.12/509/3 \\
       & 2 & $70_{-60}^{+30}$ & $2.2_{-0.3}^{+0.6}$ & ... & ... & $0.7_{-0.1}^{+0.5}$ & $1.05_{-0.02}^{+7.8}$ & 278.02/508/4 \\
        & 3 & $ \leq 1.0 \times10^{-2}$ & $2.4_{-1.0}^{+0.6}$ & $2.1_{-1.0}^{+0.8}$ & $3.1_{-2.0}^{+1.8}$ & $0.8_{-0.3}^{+0.4}$ & $1.1_{-0.2}^{+0.3}$ & 277.90/507/5 \\
    \hline
    Combined  & 1 & $ \leq 1.0 \times10^{-2}$ & $2.0_{-0.3}^{+0.4}$ & ... & ... & $1.0_{-0.3}^{+0.2}$ & $1.3_{-0.3}^{+0.3}$ & 357.73/509/3 \\
    & 2 & $60_{-50}^{+40}$ & $2.2_{-0.4}^{+0.5}$ & ... & ... & $1.0_{-0.2}^{+0.6}$ & $1.3_{-0.2}^{+0.5}$ & 357.91/508/4 \\
    & 3 & $ \leq 1.0 \times10^{-2}$ & $2.4_{-0.9}^{+0.6}$ & $1.6_{-0.6}^{+0.8}$ & $2.7_{-2.0}^{+2.1}$ & $1.0_{-0.2}^{+0.4}$ & $1.3_{-0.2}^{+0.4}$ & 357.24/507/5 \\
    \hline  
    \enddata
    \tablecomments{ Columns: (1) Observation ID number or combined; (2) Model number, where Model 1 is an absorbed redshifted power law (\texttt{phabs$\times$zphabs$\times$zpow}), Model 2 is a partially absorbed redshifted power law (\texttt{phabs$\times$(zphabs$\times$zpow$+$zpow})) with power law indices tied, and Model 3, is an absorbed redshifted broken power law (\texttt{phabs$\times$zphabs$\times$bknpower}); (3) extragalactic column density in units of $10^{22}$ cm$^{-2}$; (4) spectral index for $\Gamma_1$; (5) spectral index for $\Gamma_2$; (6) power law break energy in keV; (7) measured 2--10 keV Flux in units of $10^{-13}$ erg s$^{-1}$ cm$^{-2}$; (8) measured 2--10 keV Luminosity in units of $10^{43}$ erg s$^{-1}$; (9) the best-fit Cash statistic, degrees of freedom, and number of free model parameters. All values reported are the means from chains. The error bars reported are at the 95\% confidence level.}
    \tablenotetext{\dagger}{When fitting with the Cash statistic these values are pegged to the maximum/minimum values ($\Gamma_1 = 3$, $\Gamma_2 = 1$).}
    
    \label{tab:xraymod}
\end{deluxetable*}

We find that Observation 23328 statistically favors Model 3 over Model 1 ($\Delta C_{\mathrm{stat}}/\Delta N_{\mathrm{fp}} = 3.9$). However, the photon indices are pegged at the bounds ($\Gamma_1 = 3$, $\Gamma_2 = 1$). We therefore adopt Model 1 as the best-fit model. For Observation 23821, we find Model 1 is statistically favored. Given that Model 1 is favored by both observations and returns values that are consistent at the 95\% confidence level, we continue to fit the combined spectra. 

\begin{figure}[t]
    \centering
    \includegraphics[width=0.99\linewidth]{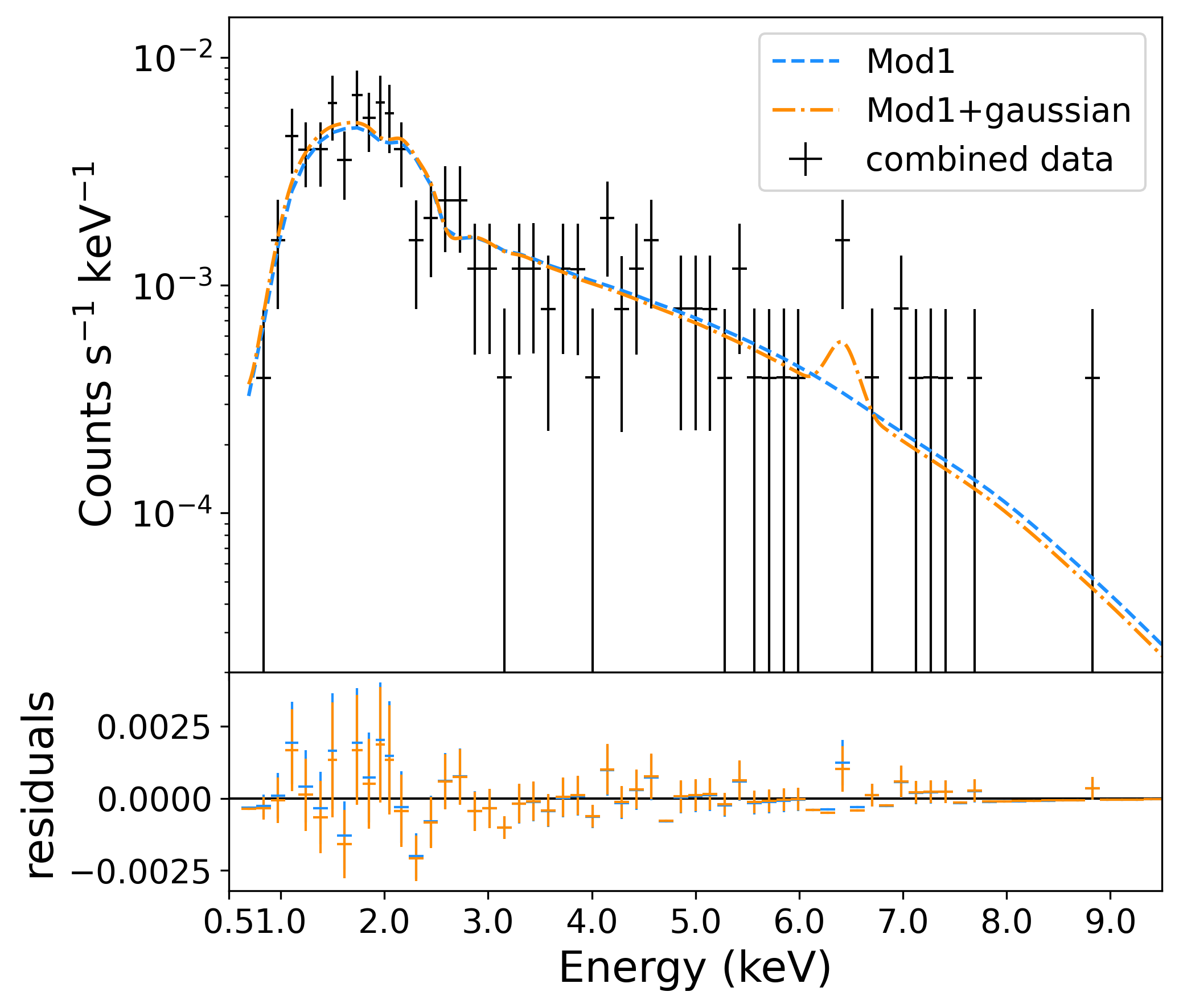}
    \caption{\textit{Top:} The absorbed 0.5 to 8 keV Chandra combined spectrum of J095036. We show the best-fit model, Model 1, a redshifted absorbed power law (\texttt{phabs$\times$zphabs$\times$zpow}; blue dashed) alongside the same model with an added Gaussian (\texttt{phabs$\times$zphabs(zpow+zgauss}); orange dash-dotted) to fit for an Fe-K$\alpha$ emission line. Although the model with added gaussian is not statistically favored we compute an equivelent width of 0.34 keV. The parameter values are shown in Table \ref{tab:xraymod}. The spectra have been rebinned for readability. \textit{Bottom:} Residuals for Model 1 (blue) and Model 1 with gaussian (orange). }
    \label{fig:xrayspec}
\end{figure}

We combine the two spectra using the CIAO \texttt{combine\_spectra} function and refit the three models to the combined spectrum. Model 1 remains the preferred fit. We measure absorbed 0.5--2 keV and 2--10 keV fluxes of $6.5^{+1.9}_{-1.0} \times10^{-14}$ erg s$^{-1}$ cm$^{-2}$ and $9.6^{+2.2}_{-3.0} \times10^{-14}$ erg s$^{-1}$ cm$^{-2}$, values in agreement with those found by \cite{breiding_chandra_2026}. This corresponds to a 0.5--2 keV luminosity of $9.0^{+2.6}_{-1.3} \times10^{42}$ erg s$^{-1}$ and a 2--10 keV luminosity of $1.2^{+0.3}_{-0.4} \times10^{43}$ erg s$^{-1}$ at $z=0.2142$. No K-corrections are applied as the values are computed directly from the spectrum. We show the combined X-ray spectrum of J095036 in Figure \ref{fig:xrayspec} along with the combined fit of Model 1. We list all key parameter values in Table \ref{tab:xraymod}, with best-fit values and confidence intervals derived from the tool \texttt{chain}.

We add a Gaussian line component to Model 1 of the combined spectrum, \texttt{(phabs $\times$ zphabs (zpow+zgaus))}, to investigate the presence of Fe K$\alpha$ emission, motivated by a modest excess near the expected rest-frame energy of 6.4 keV in the absorbed spectrum (Figure \ref{fig:xrayspec}). Given the low-count spectrum, we fix the line width to $\sigma = 0.1$ keV, consistent with the expectation of a narrow emission line, and allow only the line energy to vary. We obtain a best-fit energy of $6.6^{+0.4}_{-0.5}$ keV, consistent with neutral Fe K$\alpha$ emission. 
To assess whether the line energy is meaningfully constrained by the data, we refit the spectrum with the line energy fixed at 6.4 keV. This fixed-energy model provides the preferred fit, although the free-energy model yields a statistically consistent line energy. For the fixed-energy model, we measure an equivalent width of 0.34 keV. However, the addition of the Gaussian line component does not produce a statistically significant improvement in the fit under the Cash statistic, indicating that any Fe K$\alpha$ emission is not strongly constrained by the current data, consistent with the non-detection reported by \cite{breiding_chandra_2026}.

\section{The Multi-wavelength SED} \label{sec:SED}
In this section, we assemble a multi-wavelength SED of J095036. Following a similar analysis as presented in \cite{foord_investigating_2022} and \cite{foord_multi-wavelength_2017}, we compile all available observations of J095036 and compare them to the standard AGN SEDs presented in \cite{shang_next_2011}. K-corrections are applied using the relation for a power-law continuum established in \cite{richards_sloan_2006} given by $K(z)=-2.5(1+\alpha_\nu)\log (1+z)$ assuming $F_\nu \propto \nu^{-\alpha_\nu}$. 

Available SED photometric observations include archival data across the electromagnetic spectrum: mid-infrared from WISE (\citealt{https://doi.org/10.26131/irsa1}, \citealt{cutri_explanatory_2013}), near-infrared from 2MASS (\citealt{https://doi.org/10.26131/irsa2}, \citealt{skrutskie_two_2006}), optical from SDSS (\citealt{york_sloan_2000}), and two ultraviolet observations from GALEX (\citealt{martin_galaxy_2005}, available at MAST: \dataset[10.17909/T9H59D]{http://dx.doi.org/10.17909/T9H59D}). We correct for galactic extinction using dust maps from \cite{schlafly_measuring_2011} and reddening curves from \cite{fitzpatrick_correcting_1999}. We apply K-corrections assuming values of $\alpha_\nu=-1, -0.5,$ and $ -1.57$ for IR, optical, and UV respectively (\citealt{shang_next_2011}, \citealt{richards_sloan_2006}, \citealt{ivezic_optical_2002}). We also include our best-fit absorbed Chandra X-ray luminosities, shown as separate 0.5–2 keV and 2–10 keV data points. The luminosities for each rest-frame frequency and year for each observation are listed in Table \ref{tab:multi}.

\begin{figure*}
    \centering
    \includegraphics[width=0.9\textwidth]{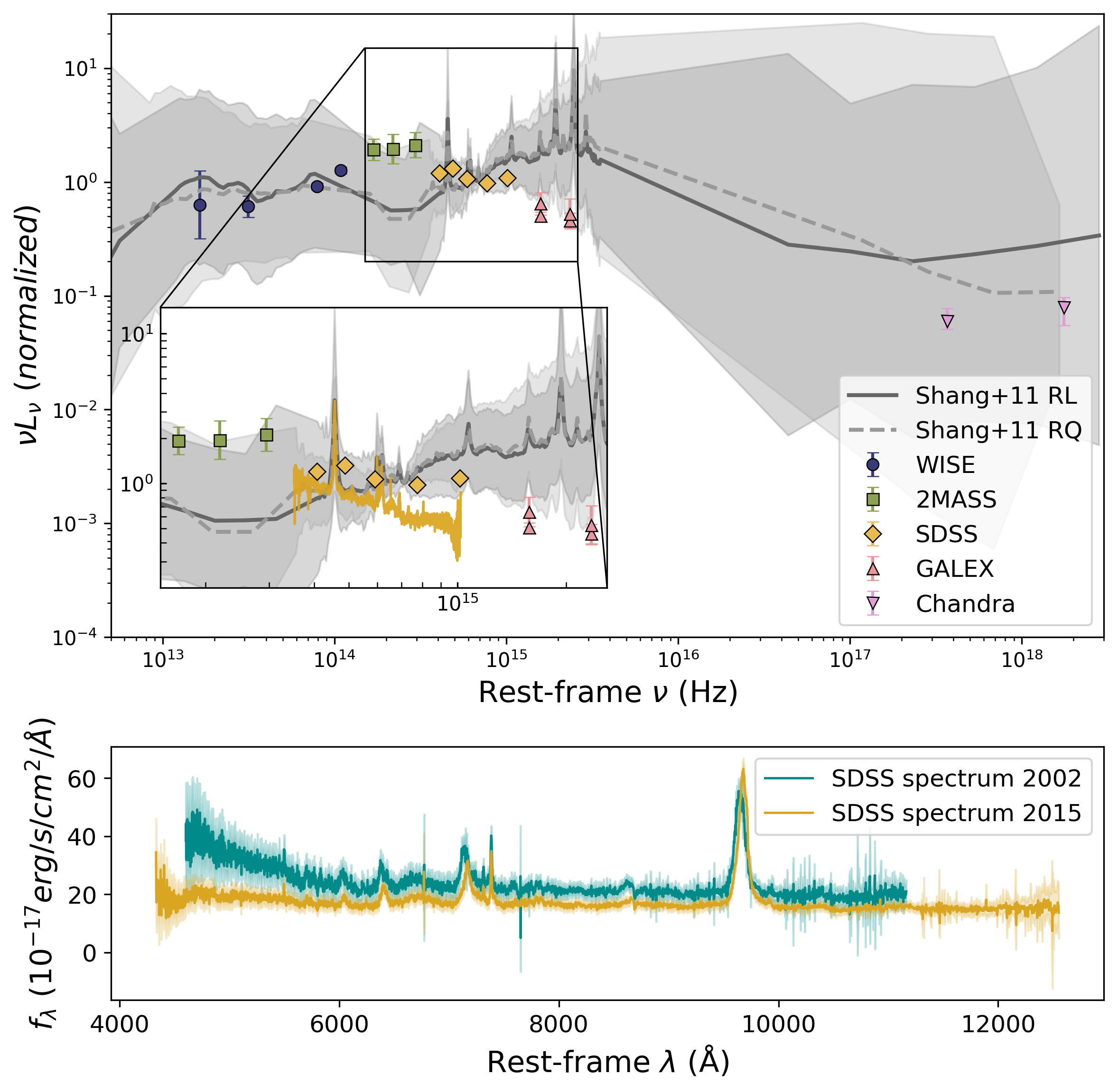}
    \caption{\textit{Top:} The rest-frame multi-wavelength SED of J095036. We plot the MIR photometry from WISE (blue circles), NIR photometry from 2MASS (green squares), optical photometry from SDSS (yellow diamonds), UV photometry from GALEX (light red triangles), and X-ray photometry from Chandra (pink downward triangles). We overplot the radio-loud (solid dark grey) and radio-quiet (dashed light gray) standard non-blazar AGN SED from \cite{shang_next_2011}. In the inset we overplot the 2015 SDSS spectrum in yellow. Error bars for the \cite{shang_next_2011} templates and archival multi-wavelength data are at the 3$\sigma$ confidence interval. 
    \textit{Bottom:} The SDSS spectra from 2002 (blue-green line) and 2015 (yellow line) in spectral flux density. Error bars are reported at a 3$\sigma$ confidence level. In general we find a good agreement between the composite quasar SED and the SED of J095036. However, J095036 appear to have a deficit in the UV emission, as well as an excess in emission from the NIR bands.}
    \label{SED}
\end{figure*}

\begin{deluxetable}{ccccc}
    \tablecaption{Multi-wavelength Luminosities}
    \tablewidth{0pt}
    \setlength\tabcolsep{5pt}
    \tablehead{
    \colhead{Survey/} & \colhead{Filter/} & \colhead{Observation} & \colhead{$\log \nu$} &\colhead{ $\log \nu L_{\nu}$} \\
    Telescope & Detector & Date & (Hz) & (erg s$^{-1}$) \\
    (1) & (2) & (3) & (4) & (5)}
    \startdata
    WISE & W4 & 2010 & $13.2$ & $43.98 \pm 0.30$ \\
    WISE & W3 & 2010 & $13.5$ & $43.96 \pm 0.09$ \\
    WISE & W2 & 2010 & $13.9$ & $44.14 \pm 0.03$ \\
    WISE & W1 & 2010 & $14.0$ & $43.98 \pm 0.03$ \\
    2MASS & K & 1999 & $14.2$ & $44.46 \pm 0.09$ \\
    2MASS & H & 1999 & $14.3$ & $44.47 \pm 0.13$ \\
    2MASS & J & 1999 & $14.5$ & $44.50 \pm 0.11$ \\
    SDSS & z & 2002 & $14.6$ & $44.26 \pm 0.03$ \\
    SDSS & i & 2002 & $14.7$ & $44.30 \pm 0.02$ \\
    SDSS & r & 2002 & $14.8$ & $44.20 \pm 0.03$ \\
    SDSS & g & 2002 & $14.9$ & $44.17 \pm 0.03$ \\
    SDSS & u & 2002 & $15.0$ & $44.21 \pm 0.03$ \\
    GALEX & NUV & 2009 & $15.2$ & $43.88 \pm 0.03$ \\
    GALEX & NUV & 2004 & $15.2$ & $43.99 \pm 0.10$ \\
    GALEX & FUV & 2009 & $15.4$ & $43.84 \pm 0.07$ \\
    GALEX & FUV & 2004 & $15.4$ & $43.90 \pm 0.13$ \\
    Chandra & ACIS-S & 2020/2021 & $17.6$ & $42.95_{-0.07}^{+0.11}$ \\
    Chandra & ACIS-S & 2020/2021 & $18.2$ & $43.07_{-0.16}^{+0.09}$ \\
    \enddata
    \tablecomments{Columns: (1) telescope or survey; (2) filter or detector; (3) year of observation; (4) rest-frame frequency at a redshift of z=0.2142 in units of Hz. The Chandra frequencies correspond to the central rest-frame frequency of the 0.5--2 keV (1.25 keV) and 2--10 keV (6 keV) range; (5) extinction and K-corrected luminosities assuming a distance luminosity of $D_L=1.07$ Gpc, in units of erg s$^{-1}$. Error bars reported for archival multi-wavelength data are at the 3$\sigma$ confidence level. See text for extinction and K-corrections applied. }
    \label{tab:multi}
\end{deluxetable}
A comparison between the multi-wavelength SED of J095036 and the radio-loud and radio-quiet \cite{shang_next_2011} single non-blazar AGN SEDs is shown in Figure \ref{SED}. We plot the multi-wavelength SED of J095036 following the normalization procedure presented by \cite{shang_next_2011}. The SED is normalized at $ \lambda \approx 4200$ \ang~ by interpolating between the rest-frame SDSS g ($\lambda \approx 3883$ \ang) and r ($\lambda \approx 5081$ \ang) bands. 

We further supplement our SED with archival SDSS spectra from 2002 and 2015. We apply extinction corrections and K-corrections to the SDSS spectra following the same methodology as the photometric data, employing \cite{schlafly_measuring_2011} dust maps and \cite{fitzpatrick_correcting_1999} reddening curves. For the 2002 spectrum, which originated from an early data release (DR1), extinction correction have already been applied and only K-corrections are necessary. The spectra from 2002 and 2015 are presented in the bottom panel of Figure \ref{SED}, with error bars indicating a 3$\sigma$ confidence level. To facilitate comparison with the broadband SED, we additionally normalize the 2015 spectrum following \cite{shang_next_2011} and overlay it on the normalized SED in the inset of Figure \ref{SED}.
 
We find that the SED of J095036 is generally consistent with the SED of \cite{shang_next_2011} across most wavelengths. However, the 2MASS near-infrared data shows a modest excess relative to the SED, albeit consistent within the 3$\sigma$ confidence interval. Given that 2MASS is the earliest dataset (observed in 1999, compared to the WISE data observed in 2010), this may indicate long-term luminosity evolution, with J095036 in a brighter state at that epoch. A similar trend is seen in the archival SDSS spectra (Figure \ref{SED}), where the 2002 spectrum is marginally more luminous than the 2015 spectrum. 

The SED also exhibits a deficit in the optical-to-UV range, beginning near the NUV band at $\sim 2600$ \AA. The 2015 SDSS spectrum similarly drops below the range covered by the \cite{shang_next_2011} template. 

\section{Results and Discussion}\label{sec:results}

In the binary SMBH scenario, the optical-to-UV deficit could arise from a gap between the mini-disks and the circumbinary disk (e.g. \citealt{roedig_evolution_2012, roedig_observational_2014, gultekin_observable_2012}) and warrants further investigation. Alternatively, the downturn may be attributed to extragalactic dust reddening in the single AGN scenario. 
In the following section, we investigate possible causes for the deviations in the SED of J095036 from the single non-blazar AGN SED from \cite{shang_next_2011}. 

\subsection{Characterizing Flux Variability and Dimming Effects} \label{sec:dimming}

Across the entire electromagnetic spectrum, we find J095036 shows evidence of both an excess and a deficit of emission with respect to a normal AGN. As shown in Figure 3, the 2MASS fluxes ($\lambda \sim 9500-19000$ \ang ; observed in 1999) appear brighter than the rest of the multi-wavelength dataset. In addition, the comparison of two SDSS spectra ($\lambda \sim 2500-8500$ \ang ; obtained in 2002 and 2015) show marginal evidence for dimming across the full optical spectral band. The UV flux also appears suppressed relative to the multi-wavelength data. Motivated by these observations, we investigate whether there is evidence for variability or dimming of J095036, given that our observations span a large range of time. In particular, previous analyses of the radio emission (observed with FIRST and VLASS) suggest the presence of a young jet that may be turning on, potentially signaling a change in the accretion state (\citealt{nyland_quasars_2020}). A more recent study of the H$\beta$ line profile finds both dimming and broadening over 22 years, although \cite{mohammed_tantalizing_2026} report a flux variability of only 20\%, which is inconsistent with large changes in the accretion-powered continuum. 

To further test for global dimming, we analyze multi-year optical data from the Zwicky Transient Facility (ZTF) obtained after 2015. The ZTF g-, r-, and i-band filters are designed to closely match those of SDSS (\citealt{bellm_zwicky_2014}), allowing for a consistent comparison and enabling a search for long-term dimming trends in the optical that would support an overall decline in the SED.

\begin{figure}
    \centering
    \includegraphics[width=\linewidth]{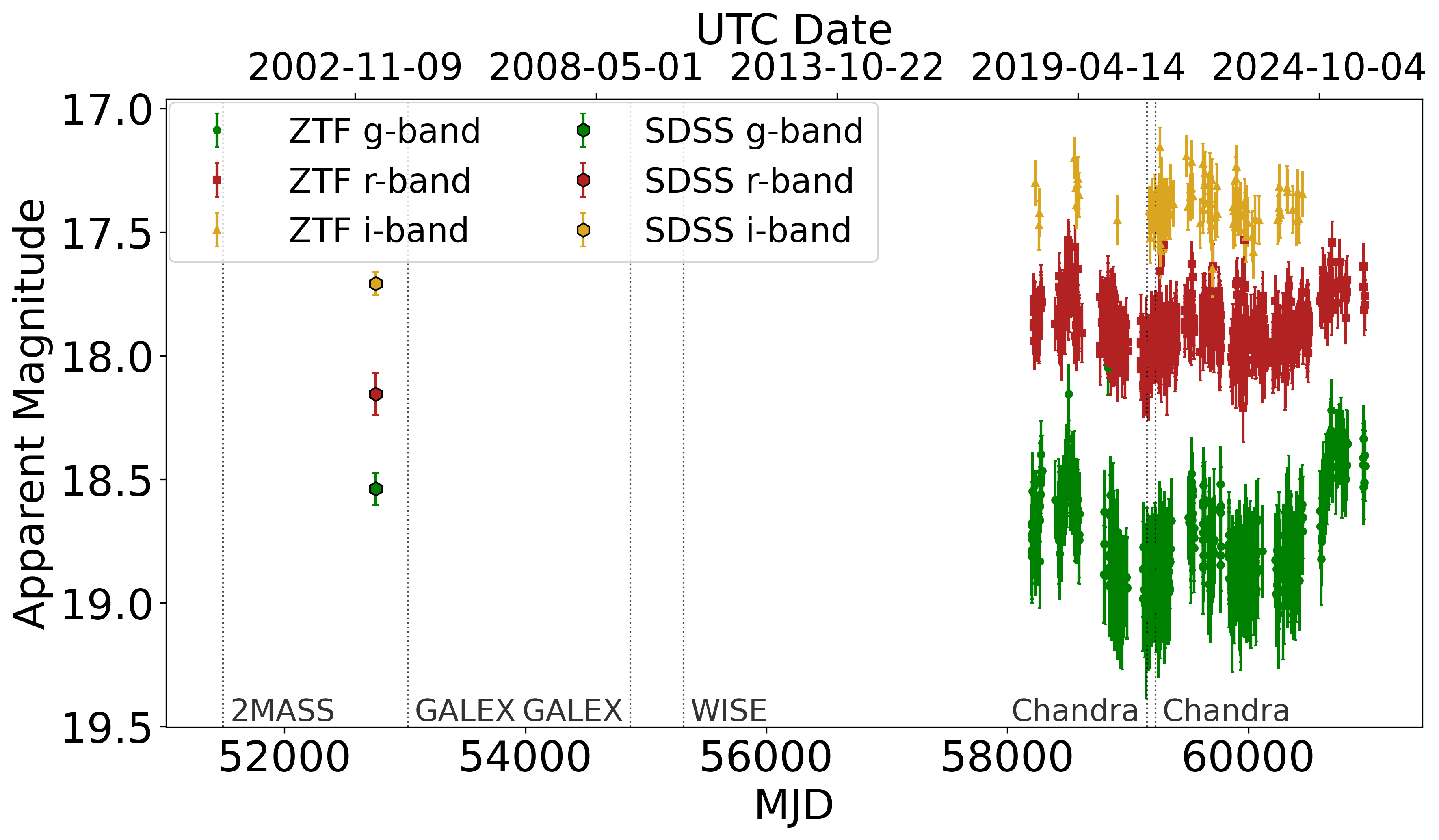}
    \caption{The comparison of light curve data from the Zwicky Transient Facility (ZTF) to their equivalent SDSS bands. G-bands are shown in green (SDSS as a hexagon, ZTF as circles), r-bands in red (SDSS as a hexagon, ZTF as squares), and i-band as yellow (SDSS as a hexagon, ZTF as triangles). Observational dates of all other SED data is shown as vertical dotted lines. No evidence of dimming can be found.}
    \label{fig:ZTF_dim}
\end{figure}

We correct all ZTF magnitudes for extinction and apply K-corrections using the same \cite{schlafly_measuring_2011} dust maps and \cite{fitzpatrick_correcting_1999} reddening curves applied to the SDSS data. We present our comparison of SDSS and ZTF magnitudes in Figure \ref{fig:ZTF_dim}. Observation dates for all other multi-wavelength data are indicated as vertical lines. We do not observe strong evidence for dimming or variability between 2002 and 2024.
 
We note that AGN variability, including dimming, in the optical bands does not necessarily translate to corresponding changes in the NIR or UV. Moreover, recent Near-Earth Object Wide-Field Infrared Survey Explorer (NEOWISE) W1 and W2 magnitudes are consistent with the earlier 2010 WISE data included in our analysis (see Figure 2 of \cite{mohammed_tantalizing_2026} where more light curves are presented), reinforcing that there is no sign of dimming in the infrared between 2010 and 2025. The lack of observed variability suggests that non-simultaneous observations are unlikely to strongly bias the mid-infrared and optical portions of the SED. Thus, the observed SED shape presented in Fig. X likely provides a reasonable representation of J095036's broadband emission. Thus, the enhanced NIR emission observed by 2MASS may suggest to atypical accretion-related processes or host galaxy contamination.

\subsection{Investigating Extragalactic Reddening} \label{sec:reddening}

A decrease in emission through the UV wavelengths is a key theoretical indication of a binary SMBH system. In typical binary SMBH systems, each black hole has a mini-disk, all embedded within a circumbinary disk (e.g. \citealt{roedig_evolution_2012}; \citealt{gultekin_observable_2012}; \citealt{tanaka_electromagnetic_2012}; \citealt{tanaka_electromagnetic_2013}; \citealt{roedig_observational_2014}). A gap forms between these disks at radii that would otherwise emit optical–UV light, leading to reduced emission in these wavelengths. The accretion structure can change significantly during the merger, influenced by binary separation and the angular momentum of incoming material.
For very close binaries or when accreting streams have low angular momentum, mini-disks may not form (\citealt{gultekin_observable_2012}; \citealt{tanaka_electromagnetic_2013}; \citealt{gold_accretion_2014}; \citealt{roedig_observational_2014}). Recent simulations, however, show mini-disks can persist at small separations but may be truncated (\citealt{bowen_relativistic_2017}; \citealt{dascoli_electromagnetic_2018}; \citealt{wang_final_2023}). If mini-disks are present, the depth of the resulting spectral “notch” depends on total mass, mass ratio, and binary separation (e.g, \citealt{roedig_evolution_2012}; \citealt{roedig_observational_2014}; \citealt{farris_binary_2014}), so the SED varies across different systems. Additionally, stream shocking can produce high-energy emission that masks the expected optical–UV drop (\citealt{roedig_observational_2014}; \citealt{farris_characteristic_2015}; \citealt{farris_binary_2015}). Previous observational studies of binary candidates have found that potential notches can be modeled by reddening of a single AGN SED (\citealt{guo_spectral_2020}, \citealt{foord_investigating_2022}).
In the following subsections, we follow a similar process as presented in \cite{foord_investigating_2022} to test whether reddening by extragalactic dust is the cause of the low UV emission present in the SED of J095036. To do so, we fit the SED with three different methods: (1) applying extinction curves presented by \cite{fitzpatrick_correcting_1999}, (2) applying the extinction model presented by \cite{goobar_low_2008}, and (3) using the SED fitting software Code Investigating Galaxy Evolution (CIGALE; \citealt{boquien_cigale_2019}). 

\subsubsection{Fitzpatrick and Goobar fitting}
To investigate possibility of reddening in the SED of J095036, we fit the SED with two analytical extinction models, following a methodology similar to that of \cite{foord_investigating_2022}. We first apply the extinction law from \cite{fitzpatrick_correcting_1999} (using the \texttt{fitzpatrick99} function from the \texttt{extinction} python package) to the quasar SED of \cite{shang_next_2011} between its defined range of rest-frame $910$ \AA\ to 6 microns (which inlcudes WISE W2 and W1, 2MASS, SDSS, and GALEX data points). We allow $A_V$ to range from 0 to 10, with an initial value of 1.5, and require $R_V$ to be positive, with an initial value of 3.1. We determine the optimal values of $A_V$ and $R_V$ by minimizing $\chi^2$ between the reddened model and the photometric data of J095036.

We then fit the extinction law presented in \cite{goobar_low_2008}. This model is defined as $A_\lambda= A_V(1-a+a(\frac{\lambda}{\lambda_V})^p)$ where $a$ is the scattering parameter and $p$ is the power-law index. $R_V$ is then computed using $R_V=1/a(0.8^p-1)$. Although originally developed for Type Ia supernovae, this model captures multiple-scattering geometries that can more strongly suppress blue emission relative to a simple foreground screen. We apply the model to all IR to UV data points (WISE, 2MASS, SDSS, and GALEX) and use a $\chi^2$ methodology to compute the reddening parameters. We allow $A_V$ to range from $0-10$, with an initial value of $1.5$; $a$ to range from 0 to 1, with an initial value of $0.6$; and $p$ to range from $-3$ to 0, with an initial value of $-1.5$.

The best-fit results for both models are summarized in Table \ref{tab:reddening}. We find best-fit values of $A_V=0.06_{-0.06}^{+0.04}$ and $A_V=0.1_{-0.1}^{+0.2}$ for Fitzpatrick and Goobar, respectively. We find that the two extinction prescriptions yield consistent and plausible parameter values within the $3\sigma$ uncertainties. The \cite{fitzpatrick_correcting_1999} model produces tighter constraints than the \cite{goobar_low_2008} formulation, with the latter driving the scattering parameter $a$ to its upper bound of 1, indicating a preference for strong multiple-scattering behavior within the allowed parameter space. The reduced $\chi^2$ values are 2.67 and 2.97 for the \cite{fitzpatrick_correcting_1999} and \cite{goobar_low_2008} models, respectively.

\begin{figure}
    \centering
    \includegraphics[width=\linewidth]{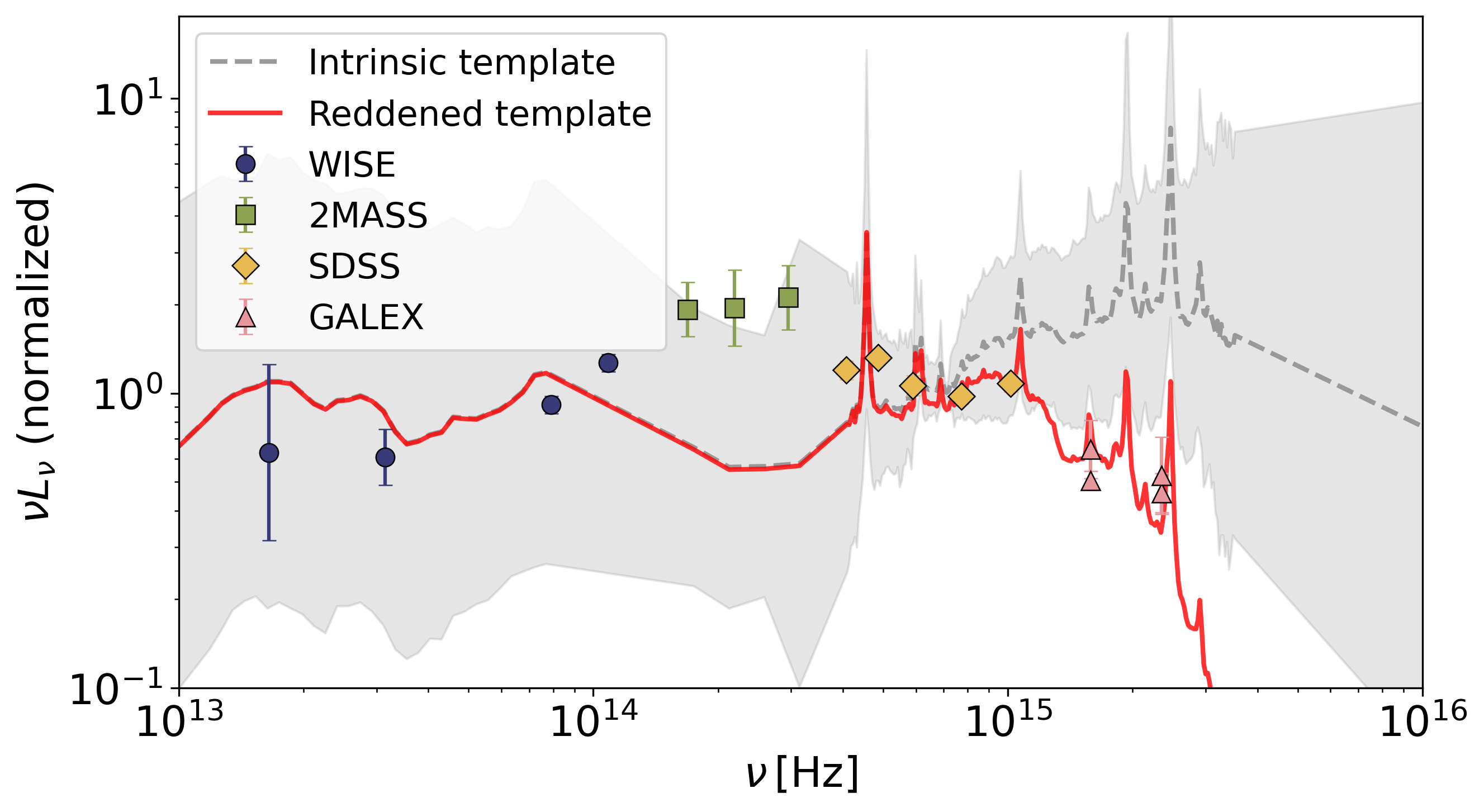}
    \caption{The best-fit reddened spectrum (red line) when fitting the radio-loud \cite{shang_next_2011} SED (grey dashed line) to the IR to UV photometric data points of J095036.  We plot the MIR photometry from WISE (blue circles), NIR photometry from 2MASS (green squares), optical photometry from SDSS (yellow diamonds), and UV photometry from GALEX (light red triangles). We apply the reddening curve from \cite{fitzpatrick_correcting_1999} and \cite{goobar_low_2008}, with the resulting line from \cite{fitzpatrick_correcting_1999} shown above, although \cite{goobar_low_2008} gives a similar result. We find best-fit values of $A_V = 0.06$ and $R_V = 0.8$ from \cite{fitzpatrick_correcting_1999}, and values of $A_V = 0.1$ and $R_V=1.9$ from \cite{goobar_low_2008} with overlapping $3\sigma$ confidence ranges. Overall, when accounting for reddening the single quasar spectrum agrees with the SED of J095036.}
    \label{fig:fitzred}
\end{figure}

The best-fit \cite{fitzpatrick_correcting_1999} model is shown in Figure \ref{fig:fitzred}; both models produce nearly identical extinction-corrected SEDs. In both cases, we find the dominant contribution to the residual $\chi^2$ arises from the near-IR 2MASS data, while the remaining photometry is well described by the reddened quasar template.

\subsubsection{CIGALE fitting}
To further investigate the effects of reddening and to better understand the potential origin of the near-infrared excess, we model our SED using the software package CIGALE (\citealt{boquien_cigale_2019}). CIGALE is a SED fitting tool that models galaxy emission from the far-UV to the radio, including X-CIGALE (\citealt{yang_x-cigale_2020}, \citealt{yang_fitting_2022}) for the inclusion of X-rays, to derive physical properties such as star formation rate, stellar mass, and dust attenuation. CIGALE has been successfully used to characterize AGN in merging galaxies (e.g., \citealt{marca_dust_2024}; \citealt{troncoso_bass_2025}), making it a suitable framework for modeling the broadband emission of J095036. However, if J095036 is a true binary SMBH system, it would represent a later stage of the merger process than the systems typically studied in these works. We adopt a delayed star formation history model with an optional exponential burst to generate the grid of models. Additionally, we use \cite{bruzual_stellar_2003} single stellar population models using the initial mass function from \cite{chabrier_galactic_2003}. We include the standard nebular emission generated by \cite{inoue_rest-frame_2011}, and the modified starburst dust attenuation model using \cite{calzetti_dust_2000} attenuation curves. The dust emission is modeled following \cite{dale_two-parameter_2014} and AGN emission is modeled using skirtor models (\citealt{stalevski_3d_2012}, \citealt{stalevski_dust_2016}). X-ray emission is fit using models from \cite{yang_x-cigale_2020}. A table of the input parameter values is shown in Table \ref{tab:reddening}. This combination of input parameters, alongside the default values not presented in Table \ref{tab:reddening}, produces 362,880 different fits. Our initial analysis, using all the data points presented in Table \ref{tab:multi}, yields a best-fit reduced $\chi^2$ value of 3.80. The CIGALE software provides both the minimum reduced $\chi^2$ solution (hereafter the best-fit model) and posterior probability distributions with associated uncertainties for each parameter.

To assess the origin of the elevated $\chi^2$ value and determine whether specific wavelength regions drive the tension with the model, we compute the per-band contributions to $\chi^2$ using the \texttt{save$\_$chi2} flux outputs. This step is motivated by the need to identify whether the discrepancy is driven by isolated outliers or by systematic mismatches in particular spectral regions, which may indicate additional complexity in the accretion environment beyond that captured by the single-AGN SED model. 

We find the largest residuals in the 2MASS bands, WISE W1 and W3, and SDSS u band, each contributing $\chi^2 > 5$. All three 2MASS points show significant positive residuals relative to the model, consistent with our qualitative comparison to \cite{shang_next_2011} in Section \ref{sec:SED}. We remove the 2MASS measurements and refit the model to evaluate whether the remaining SED can be better described under the same attenuation assumptions.

\begin{figure*}
    \centering
    \includegraphics[width=0.8\linewidth]{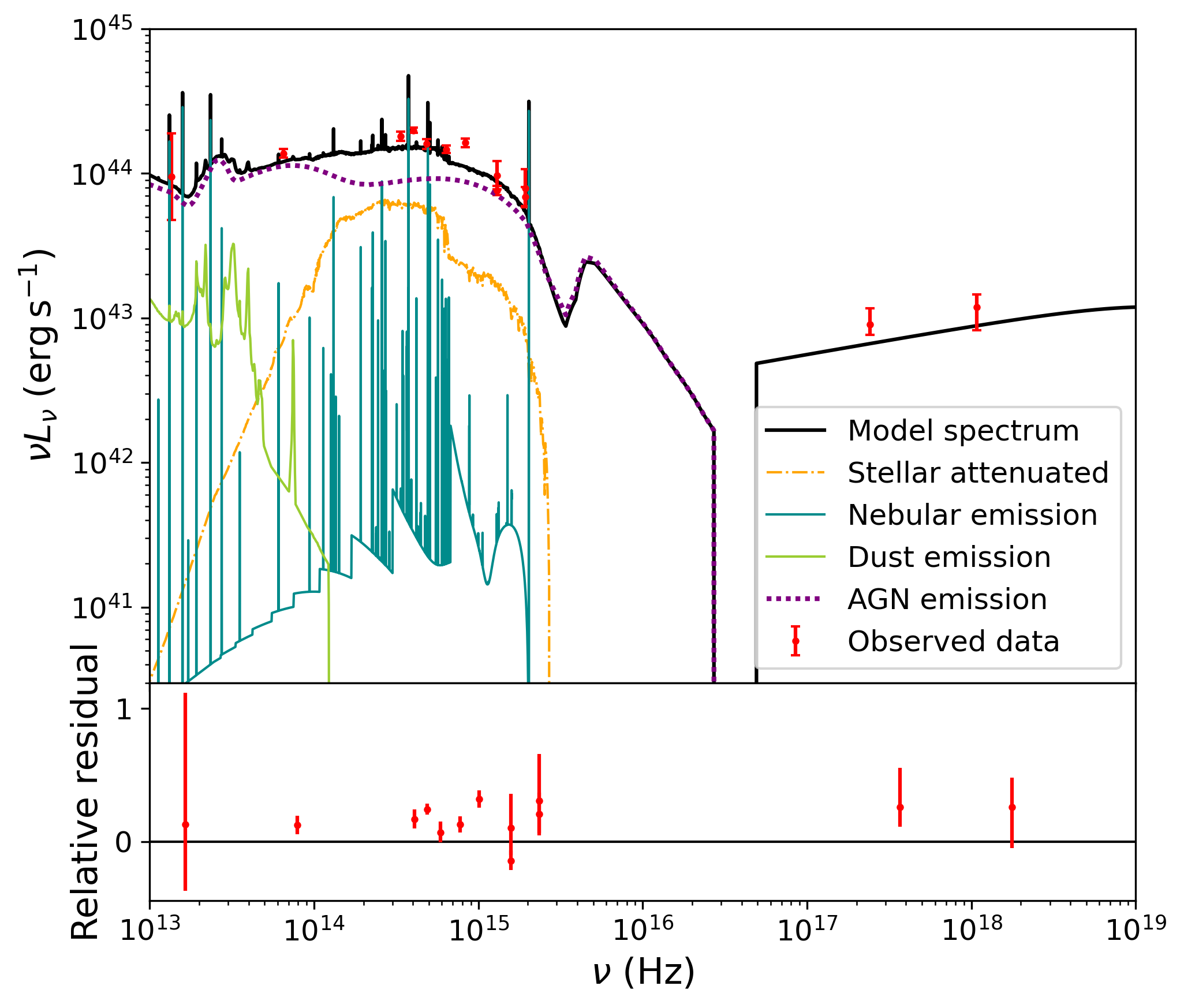}
    \caption{\textit{Top:} The CIGALE Best-fit Model SED. Photometric data for J095036 is shown as red points, with 2MASS and WISE W1 and W3 removed as was done in the fitting. We plot the total CIGALE model spectrum (black line), the stellar attenuated emission (yellow dash-dotted line), the nebular emission(blue-green line), the dust emission (light green line), and the AGN emission(purple dotted line). This model produces a reduced $\chi^2=1.00$. \textit{Bottom:} The relative residual between the total CIGALE model spectrum and the SED data of J095036 defined as (obs-model)$/$obs.}
    \label{fig:CIGALE}
\end{figure*}

To assess the origin of the elevated $\chi^2$ value and determine whether specific wavelength regions drive the tension with the model, we compute the per-band contributions to $\chi^2$ using the \texttt{save$\_$chi2} flux outputs. This step is motivated by the need to identify whether the discrepancy is driven by isolated outliers or by systematic mismatches in particular spectral regions, which may indicate additional complexity in the accretion environment beyond that captured by the single-AGN SED model. 
After excluding the 2MASS data, the total reduced $\chi^2$ decreases to 2.16, but WISE W1, W3, and the SDSS u band remain the dominant contributors to the elevated $\chi^2$. At this stage, we consider whether these remaining tensions reflect limitations of the attenuation model in capturing potential optical-UV suppression, or whether they indicate additional wavelength-dependent variability not accounted for in a static SED framework. We then repeat the fit excluding WISE W1 and W3 in addition to the 2MASS bands, yielding a best-fit $\chi^2 = 1.00$. The resulting parameter values are reported in Table \ref{tab:reddening}, with uncertainties derived from the posterior distributions. The best-fit model and SED is shown in Figure \ref{fig:CIGALE}.

Across all iterations, the inferred physical parameters remain consistent within $3\sigma$, indicating that the removal of these bands does not significantly bias the inferred solution. The star formation rate (SFR) shows a mild decreasing trend in successive fits, which is to be expected since the removed bands are dominated by stellar emission, but remains statistically consistent within uncertainties. CIGALE finds consistent, physically reasonable values of $A_V \sim0.2-0.3$.

Although the resulting parameters remain consistent within $3\sigma$, the direction of the changes provides information on the origin of the residuals. Excluding the excess infrared bands results in a modest decrease in both stellar mass (full: $2.1 \times 10^{10} M_{\odot}$, 2MASS removed: $1.6 \times 10^{10} M_{\odot}$, final: $1.5 \times 10^{10} M_{\odot}$) and SFR (full: $5.3\ M_{\odot}$ yr$^{-1}$, 2MASS removed: $4.1\ M_{\odot}$ yr$^{-1}$, final: $3.3\ M_{\odot}$ yr$^{-1}$), accompanied by an increase in the inferred AGN infrared fraction ($f_{AGN_{IR}}$; full: $0.6$, 2MASS removed: $0.7$, final: $0.7$). These results are expected, as the near-infrared 2MASS bands primarily trace the underlying stellar continuum, while the mid-infrared WISE bands are dominated by emission from AGN-heated dust. When the 2MASS, W1, and W3 bands are excluded, the inferred stellar mass, star formation rate, and IR AGN fraction all shift as the CIGALE model adjusts to reproduce the remaining photometric constraints. However, the magnitude of these changes remains within the uncertainties, indicating that the overall physical interpretation is robust.

Because the 2MASS bands show large positive residuals in the CIGALE fit, we investigate whether the excess near-infrared emission could be attributed to unresolved host-galaxy contamination. We independently estimate the host-galaxy contribution using the empirical relation from \cite{jalan_empirical_2023} and compare these values with the AGN/host decomposition inferred from CIGALE. Using quasar spectra from SDSS, \cite{jalan_empirical_2023} calibrate an empirical relation between the host-galaxy fraction and the AGN luminosity at $5100\ $AA\, finding hosting fractions consistent with those derived from image decomposition methods. We estimate the $5100\ $AA\ luminosity of J095036 via interpolation using the available photometry as well as the two SDSS spectra. This yields luminosities (and host-galaxy fractions) of
$1.6 \times 10^{44}$ erg s$^{-1}$ ($38_{-1}^{+2}$\%), $3.2 \times 10^{44}$ erg s$^{-1}$ ($35_{-1}^{+2}$\%), and $2.3 \times 10^{44}$ erg s$^{-1}$ ($36_{-1}^{+2}$\%) for the photometric SED, 2002 spectrum, and 2015 spectrum, respectively. These estimates are consistent with the host fraction inferred from CIGALE ($\approx30\%$), calculated as $(1-f_{AGN_{IR}})\times100$. We note, however, that these two quantities are not strictly equivalent, as the empirical relation constrains the host contribution near $5100\ $AA\, whereas CIGALE describes the AGN contributed integrated over the infrared. Nevertheless, the agreement suggests that the near-infrared excess is unlikely to be caused solely by an underestimated host-galaxy contribution, and may reflect limitations in the SED templates, or additional complexity in the AGN emission.

Although the host-galaxy fraction inferred from CIGALE is consistent with an independent estimate, residual discrepancies could still arise from limitations in the assumed stellar or AGN templates. To further investigate whether the infrared residuals reflect an incorrect spectral shape rather than  an incorrect normalization, we compare the observed and modeled broadband colors (using the full data set model). The (J-K) color of J095036 (0.5 mag) is well reproduced by the model (0.5 mag), suggesting that the stellar continuum shape in the near-infrared is adequately captured, despite the elevated 2MASS normalization indicated by the large positive residuals. 
In contrast, the model predicts redder (W1-W2) and (W2-W3) colors ($0.21$ mag and $1.1$ mag) than observed ($-0.01$ mag and $0.6$ mag), suggesting that the adopted AGN/dust emission templates may not fully capture the observed mid-infrared emission.

We test how our CIGALE fit is affected by the stellar population synthesis model or AGN dust geometry model by repeating our full analysis (all iterations), but using the updated \cite{bruzual_stellar_2003} (2019; CB19) stellar population models and the \cite{fritz_revisiting_2006} AGN torus prescription. CB19 incorporates modern stellar evolution tracks and empirical libraries with finer grids and uses updated prescriptions for hot stars as compared to BC03. The main difference between the SKIRTOR and Fritz AGN torus models lies in the torus structure. Fritz uses a smooth distribution and SKIRTOR a clumpy two-phase distribution with high-density clumps embedded in a lower-density medium. The change from BC03 to CB19 would specifically impact the NIR wavelength regime as it traces the stellar continuum, but we find this change has little effect on any of the inferred parameters. The \cite{fritz_revisiting_2006} torus model provides a lower reduced $\chi^2$ value for the full initial model ($\chi^2=2.4$), but produces redder infrared colors than the data for all 3 colors (J-K: $0.7$ , W1-W2: $0.2$, W2-W3: $0.7$), indicating a poorer match to the observed NIR-to-MIR spectral shape despite the improved global statistic. These tests suggest that the residual infrared mismatch is not resolved by alternative standard SED prescriptions and likely reflects limitations in the adopted templates or additional complexity in the infrared emission that the models cannot capture.

In comparing our best-fit CIGALE outputs with the \cite{fitzpatrick_correcting_1999} and \cite{goobar_low_2008} models, we find good agreement in $A_V$ (see Table \ref{tab:reddening}). Given these results, we find that the SED of J095036, excluding its near-IR behavior, can be well described as a single reddened AGN.

\subsection{Comparison to Other Observational Analyses}
We compare the derived spectral properties of J095036 with the quasar SED analysis of \cite{lusso_nature_2014}. \cite{lusso_nature_2014} analyze the SEDs of 32 type-1 AGN identified as binary candidates due to high velocity offsets between their broad and narrow emission lines \citep{tsalmantza_systematic_2011} . \cite{lusso_nature_2014} investigate whether multi-wavelength SED properties can provide additional evidence for binary SMBH candidates, finding enhanced $12 \mu m$ emission relative to their control sample. To compare J095036 with their sample,  we calculate the optical and near-infrared spectral slopes using the same wavelength intervals adopted by \cite{lusso_nature_2014}. Using our best-fit CIGALE model, we find the resulting optical  ($\alpha_{\rm opt}=-0.51$) and near-infrared ($\alpha_{\rm NIR}=0.29$) slopes are broadly consistent with the range of values observed for both the binary candidate and control samples. However, when using the observed photometric data, we obtain redder slopes, ($\alpha_{\rm opt}\approx-0.96$ and $\alpha_{\rm NIR}\approx0.51$) although these values remain consistent with the range observed in their sample (see Figure 2 of \cite{lusso_nature_2014} for comparison). The difference between the template-derived and photometry-derived slopes reflect the enhanced infrared emission relative to the standard AGN SED identified throughout our analysis.

We additionally calculate the $12 \mu m$ rest-frame luminosity and $12 \mu m$ emission size of J095036 to compare to the results presented in \cite{lusso_nature_2014}. \cite{lusso_nature_2014} assume a linear relation and use $L_{12\mu}$ to calculate the emission size for both the binary and single AGN samples in order to identify other regions in parameter space where binaries differ from single AGN. We estimate an inferred $12 \mu m$ rest-frame luminosity of $L_{12\mu{\rm m}}=1.27\times10^{44}\ erg$ $s^{-1}$ and characteristic infrared emitter size of $s_{12}=6.4$ pc. These values are constant with both the binary candidate sample and the control sample. 

Our MW SED also provides the opportunity to evaluate for a recoiling black hole scenario. As discussed by \cite{lusso_nature_2014}, a recoiling SMBH may leave behind the dusty torus, resulting in a suppressed mid-infrared emission component. However, J095036 exhibits strong $12\mu m$ emission and a compact inferred infrared-emitting region, consistent with the presence of a nuclear dust component. Combined with the velocity evolution arguments presented by \cite{mohammed_tantalizing_2026}, the SED properties of J095036 disfavor the recoiling SMBH scenario.

We emphasize that our findings that the MW SED of J095036 is well-described by a reddened AGN does not exclude the binary AGN hypothesis. In particular, the persistent near-IR excess may indicate an additional emission component not captured by a simple dust screen, such as enhanced hot circumnuclear dust emission, multiple unresolved nuclear components, or complex merger-driven obscuration geometry (e.g, \citealt{kishimoto_dust-eliminated_2005}, \citealt{hopkins_origins_2012}, \citealt{ricci_growing_2017}, \citealt{ricci_hard_2021}, \citealt{barrows_census_2023}). This excess remains a key feature distinguishing the source from typical single-AGN SEDs.

\section{Conclusions}\label{sec:conclusion}
In this work, we present a multi-wavelength analysis of binary SMBH candidate SDSS J095036.75+512838.1 (J095036). J095036 was first identified in a search for SDSS quasars exhibiting broad H$\beta$ line shifts $>1000$ km s$^{-1}$ by \cite{eracleous_large_2012}, and subsequently flagged as a strong binary candidate based on systematic velocity shifts of more than 2000 km s$^{-1}$(\citealt{runnoe_large_2017},\citealt{mohammed_tantalizing_2026}). To search for evidence of two accreting SMBHs, we construct a broadband SED incorporating recent Chandra X-ray observations alongside archival MIR (WISE), NIR (2MASS), optical (SDSS), and UV (GALEX) data. The principal findings and implications of this work are:
\begin{enumerate}
    \item We analyze Chandra X-ray observations (Obs IDs 23328 and 23821) of J095036 by fitting the 0.5-8 keV spectra from each observation with 3 different models to search for evidence of a binary. We find the simple absorbed power law is the best-fit model for both observations and the combined data. We add a Gaussian component to model the Fe K$\alpha$ line and derive an upper limit, obtaining an equivalent width of 0.34 keV. We measure the absorbed 0.5--2 keV and 2--10 keV fluxes of $6.5^{+1.9}_{-1.0} \times10^{-14}$ erg s$^{-1}$ cm$^{-2}$ and $9.6^{+2.2}_{-3.0} \times10^{-14}$ erg s$^{-1}$ cm$^{-2}$. This corresponds to a 0.5--2 keV luminosity of $9.0^{+2.6}_{-1.3} \times10^{42}$ erg s$^{-1}$ and a 2--10 keV luminosity of $1.2^{+0.3}_{-0.4} \times10^{43}$ erg s$^{-1}$ at z=0.2142, with error bars at the 95\% confidence level.
    \item We combine all available multi-wavelength observations and compare the SED to a standard non-blazar AGN SED. The available SED observations include MIR photometry from WISE, NIR photometry from 2MASS, optical photometry from SDSS, and UV photometry from GALEX. We find good agreement between the SED of J095036 and that of a standard AGN; however, we find a deficit of emission from SDSS J095036 at UV frequencies, and an excess emission in the NIR.
 
    \item We investigate whether discrepancies between the multi-epoch observational data and the AGN SED model can be driven by dimming or accretion-state variability. We compare ZTF light-curve measurements with SDSS photometry, providing a temporal baseline of 22 years, and find no evidence of significant optical variability or sustained dimming over this period. However, the absence of optical dimming, lack of measured MIR variability, does not rule out variability at near-infrared wavelengths, and changes in the near-infrared emission remain a viable possibility. 
    \item We investigate if reddening can model the deficit of emission at $\sim2600$ \ang. Following the procedure of \cite{foord_investigating_2022}, we apply analytical extinction models from \cite{fitzpatrick_correcting_1999} and \cite{goobar_low_2008} to a standard non-blazar AGN SED to fit the photometric data points for J095036. We find that both models can well-describe the observed emission, with best-fit values of $A_V=0.06$ and $R_V=0.8$ from \cite{fitzpatrick_correcting_1999}, and $A_V=0.1$ and $R_V=1.9$ from \cite{goobar_low_2008}. 
    
 \item We fit the entire multi-wavelength SED using CIGALE, finding a best-fit model once removing high $\chi^2$ points (2MASS, W1, and W3). Our best-fit model yields a $\chi^2=1.00$ and $A_V=0.3$, consistent with our analytical analysis.  We test whether the the near-IR residuals are affected by the stellar population synthesis and AGN dust geometry models, and find that the residuals can not be resolved by alternative SED prescriptions. The excess of emission in the near-IR is may be due to additional complexity in J095036 that current models can not capture.
 
\end{enumerate}

Our analysis indicates that J095036 is best explained as a single reddened AGN system. This does not preclude that the reddened AGN is a member of a bound binary. However, the scarcity of confirmed binary SMBHs means that diagnostic methods for distinguishing binaries from single AGNs are still limited. These challenges are compounded by the dependence of circumbinary accretion signatures on the specific physical parameters of each binary SMBH system. For J095036, high-resolution near-infrared observations would enable a clearer separation of the AGN and host-galaxy contributions, helping to determine the origin of the infrared excess and resolve the current discrepancies in the SED. Although the SED is constructed from non-simultaneous observations and may not fully capture the instantaneous spectral shape of J095036, the lack of significant variability suggests that the overall luminosity scale remains relatively stable across the different epochs.

Lastly, targeted follow-up observations of the Fe K$\alpha$ emission line with higher spectral resolution could provide valuable constraints on the nature of the X-ray emitting region and on the presence of possibly binary SMBH signatures.
\begin{table*}
    \caption{Reddening Fitting Parameters}
    \centering
\label{tab:reddening}
\setlength\tabcolsep{5pt}
\begin{tabular*}{0.75\textwidth}{ccc}
    \hline
	\hline
	\multicolumn{3}{c}{CIGALE Input Parameters} \\
	\hline \\ [-2.7ex]
	\multicolumn{1}{c}{Module} & \multicolumn{1}{c}{Parameter} & 
    \multicolumn{1}{c}{Value}  \\
	\multicolumn{1}{c}{(1)} & \multicolumn{1}{c}{(2)} & \multicolumn{1}{c}{(3)} \\ [1.ex]
	\hline \\ [-2.7ex]
    duststatt modified starburst& EBV lines & 0.05, 0.1, 0.15, 0.2, 0.25, 0.3 \\ [0.4ex]
     & powerlaw slope & -1.0, -0.8, -0.6, -0.4, -0.2, 0.0, 0.2 \\
    \hline
     skirtor2016 & inclination & 0, 10, 20, 30 \\
     & delta & -0.2, -0.1, 0.0, 0.1, 0.2 \\
     & fracAGN & 0.3, 0.4, 0.5, 0.6, 0.7, 0.8, 0.9, 0.9999 \\
     & EBV & 0.0, 0.01, 0.05, 0.1, 0.15, 0.2 \\
    \hline
    yang20 & max alpha ox & 1 \\
    \hline
    \hline
    \multicolumn{3}{c}{CIGALE Output Parameters} \\
	\hline \\ [-2.7ex]
	\multicolumn{1}{c}{Data run} &
	\multicolumn{1}{c}{Parameter} &
	\multicolumn{1}{c}{Value} \\
	\multicolumn{1}{c}{(1)} & \multicolumn{1}{c}{(2)} & \multicolumn{1}{c}{(3)} \\ [1.ex]
	\hline \\ [-2.7ex]
    all data & $A_B$ & $0.4\pm0.2$  \\
      $\chi^2=3.80$& $A_V$ & $0.3 \pm 0.2$ \\
     & $f_{AGN_{IR}} $ & $0.6 \pm 0.1$ \\
     & $log(M_*)$ & $10.3 \pm 0.2$ \\
     & SFR & $5.3 \pm 1.8$\\
     & $\alpha_{ox}$ & $-1.50 \pm 0.05$ \\
     \hline
     no 2MASS & $A_B$ & $0.3 \pm 0.2$  \\
     $\chi^2=2.16$& $A_V$ & $0.2 \pm 0.2$ \\
     & $f_{AGN_{IR}} $ & $0.7 \pm 0.1$ \\
     & $log(M_*)$ & $10.2 \pm 0.2$ \\
     & SFR & $4.1 \pm 1.6$ \\
     & $\alpha_{ox}$ & $-1.50 \pm 0.02$ \\ 
     \hline
     no 2MASS, W1, W3 & $A_B$ & $0.4 \pm 0.2$  \\
     $\chi^2=1.00$& $A_V$ & $0.3 \pm 0.2$  \\
     & $f_{AGN_{IR}} $ & $0.7 \pm 0.2$ \\
     & $log(M_*)$ & $10.2 \pm 0.2$ \\
     & SFR & $3.3 \pm 1.8$ \\
     & $\alpha_{ox}$ & $-1.50 \pm 0.04$ \\
     \hline
     \hline
    \multicolumn{3}{c}{Fitzpatrick and Goobar Output Parameters} \\
	\hline \\ [-2.7ex]
	\multicolumn{1}{c}{Model} &
	\multicolumn{1}{c}{Parameter} &
	\multicolumn{1}{c}{Value} \\
	\multicolumn{1}{c}{(1)} & \multicolumn{1}{c}{(2)} & \multicolumn{1}{c}{(3)} \\ [1.ex]
	\hline \\ [-2.7ex]
    \cite{fitzpatrick_correcting_1999}& $A_V$ & $0.06_{-0.06}^{+0.04}$ \\
      $\chi^2=2.67$& $R_V$ & $0.8_{-0.8}^{+1.0}$ \\
    \hline
    \cite{goobar_low_2008}& $A_V$ & $0.1_{-0.1}^{+0.2}$ \\
     $\chi^2=2.97$ & a & $1.0_{-0.8}^{+0.0}$ \\
     & p & $-1.9_{-1.1}^{+0.7}$ \\
     & $R_V$ & $1.9_{-0.8}^{+11.5}$ \\
     \hline
\end{tabular*}

\textit{Top:} Columns: (1) CIGALE module; (2) CIGALE parameter; (3) input values for each parameter. All other inputs not listed are at the CIGALE default value.
\textit{Middle:} Columns: (1) Data run defined by which SED values were included in the CIGALE fit; (2) CIGALE parameter; (3) parameter value from the best-fit $\chi^2$ run with error bars calculated from the posterior distribution to $2\sigma$.  
\textit{Bottom:} Columns: (1) Reddening model;  (2) fitted parameter; (3) best-fit parameter value through $\chi^2$ fitting. Error bars reported to $3 \sigma$.
\end{table*}
\bibliography{ref}
\bibliographystyle{aasjournalv7}
\end{document}